\documentclass[]{iopart}
\usepackage{graphicx, epsfig, iopams}
\usepackage[colorlinks, urlcolor=blue, citecolor=blue]{hyperref}

\def\be{\begin{equation}}
\def\ee{\end{equation}}
\def\bea{\begin{eqnarray}}
\def\eea{\end{eqnarray}}

\begin{document} 

\title[Microstructure of hydrogenated amorphous silicon]{Microstructure, vacancies 
and moments of nuclear magnetic resonance of hydrogenated amorphous silicon} 

\author{Parthapratim Biswas\footnote{Author to whom any correspondence should be addressed.} and Rajendra Timilsina}
\address{Department of Physics and Astronomy, The University of Southern Mississippi, Hattiesburg, MS 39406, USA}
\eads{{\mailto{Partha.Biswas@usm.edu}, \mailto{Rajendra.Timilsina@eagles.usm.edu}}}


\begin{abstract}
Recent experiments on hydrogenated amorphous silicon using infrared absorption spectroscopy 
have indicated the presence of mono- and divacancy in samples for concentration of up to 14\% 
hydrogen. Motivated by this observation, we study the microstructure of hydrogen in two model 
networks of hydrogen-rich amorphous silicon with particular emphasis on the nature of the 
distribution (of hydrogen), the presence of defects, and the characteristic features of the 
nuclear magnetic resonance spectra at low and high concentration of hydrogen. Our study 
reveals the presence of vacancies, which are the built-in features of the model networks. 
The study also confirms the presence of various hydride configurations in the networks that 
include from silicon monohydrides and dihydrides to open chain-like structures, which have 
been observed in the infrared and nuclear magnetic resonance experiments. The broad and the 
narrow line widths of the nuclear magnetic resonance spectra are calculated from a knowledge 
of the distribution of spins (hydrogen) in the networks.  
\end{abstract}



\section{Introduction }
Since its first preparation in 1969 using glow discharge deposition technique~\cite{Chittick}, hydrogenated 
amorphous silicon is possibly the most extensively studied material of technological 
importance. The material is widely used in solar cells~\cite{Carlson}, thin film transistors~\cite{Snell,Yaniv,Hack}, 
memory switching circuits~\cite{Ovshinsky,Owen}, photosensors~\cite{Takeda}, and numerous other 
electronic devices~\cite {Street}. A multitude of experimental data are available that 
address almost every aspect of structural, electronic, optical, and vibrational properties 
of the material~\cite{Street, Morigaki}. Fundamental to the understanding of the Staebler-Wronski 
(SW)~\cite{Staebler} effect---the degradation of the material upon prolong exposure to light 
irradiation---is the distribution and the dynamics of hydrogen atoms in amorphous silicon networks. 
Since many of the proposed microscopic mechanisms that attempt to explain the SW effect are based 
on bond-breaking models~\cite{sw-model1, sw-model2}, it is essential to understand the 
local environment of hydrogen atoms in amorphous silicon networks. While there exist a number of 
experimental studies that address the distribution of hydrogen in amorphous silicon samples via 
nuclear magnetic resonance (NMR)~\cite{Gleason, Reimer, Baum, Rutland, Kuo, Carlos} and infrared 
(IR) spectroscopy~\cite{Gleason, Reimer, ir, Schropp}, there are very few theoretical 
studies~\cite{Chak, Drabold, Raj, Lee, Kim, Zhang} that address the problem explicitly.  

In this paper we study the hydrogen microstructure of realistic models of 
hydrogen-rich amorphous silicon at low and high concentration. Since a set of one-dimensional 
NMR data cannot be uniquely mapped onto a three-dimensional real space distribution of 
hydrogen without further information and assumptions, it is more appropriate and useful to 
address the problem by employing suitable models of hydrogenated amorphous silicon. 
The microstructure of the model networks has been studied with particular emphasis on 
the characteristics of the real space distribution of hydrogen in the networks, and has been 
compared directly to the results obtained from the NMR and IR experiments. 
A theoretical estimate of the width of the resonance curve has been obtained from a 
knowledge of the position of the spins in the network using a suitable approximation to the 
shape of the NMR line spectra. The approach provides a direct route to study the distribution 
of hydrogen, the presence of different hydride configurations, and the size of the various 
hydrogen clusters in the networks.  The evolution of the microstructure with the concentration 
of hydrogen in the networks has been also addressed, and a comparison has been made to the 
experimental data.

The plan of the paper is as follows. In section 2, we briefly review the results 
from the earlier theoretical and experimental studies with particular emphasis 
on the results obtained from the NMR and IR spectroscopies of hydrogenated amorphous silicon. 
This is followed by a discussion of the microstructure that we have 
observed in the model networks at low and high concentration in section 3.  
Section 4 addresses the presence of vacancies in the model networks and 
compares the results to all available experimental data. In section 5 we address how 
to calculate the line widths of the NMR spectra from a knowledge of the distribution of 
the spins in amorphous networks. This is followed by a conclusion of our work in section 6.


\section{Earlier works on hydrogen microstructure: A brief review}
Nuclear magnetic resonance (NMR)~\cite{Gleason, Reimer, Baum, Rutland, Kuo, Carlos} and 
infrared (IR) spectroscopy~\cite{Gleason, Reimer, ir, Schropp} are the two principal 
experimental techniques that can address hydrogen microstructure in amorphous silicon. 
The former provides useful information about the nature of the distribution of hydrogen 
in amorphous networks via dipolar interaction between the spins in hydrogen atoms, whereas the latter 
identifies the presence of various hydrogen bonding configurations by probing 
the bonding environment of hydrogen atoms. 
Among the early works, the proton magnetic resonance study by Reimer \etal\cite{Reimer} 
on plasma-deposited samples is the first to indicate the presence of 
inhomogeneity in the hydrogen distribution in amorphous silicon. The study also revealed
that hydrogen atoms could reside in amorphous networks as small clusters and in a
dilute environment. While NMR studies cannot provide directly a description of the 
three-dimensional spin distribution in the network, it is possible to 
infer useful information about the size of hydrogen clusters and the nature of the 
distribution (either sparse or dense) of hydrogen atoms in the samples by analyzing the 
shape and the width of an NMR spectrum. A typical NMR spectrum of a device-quality sample 
shows the presence of both narrow and broad line widths, which can be approximated as a 
convolution of a truncated Lorentzian and a Gaussian distribution. 
An analysis of the NMR spectra of amorphous silicon samples with 8--32 at.\% H by 
Reimer \etal\cite{Reimer} indicated that the broad line width of the spectrum lied in the 
range 22--27 kHz, whereas the narrow line width was of the order of 3--5 kHz. Similar observation 
was also reported in the NMR experiments by Carlos \etal\cite{Carlos}, who also noted 
a variation of the line width of up to 20\% depending on the nature of amorphous samples 
used in the measurements.

A definitive picture about the microstructure emerged with the advent of multiple-quantum 
nuclear magnetic resonance (MQ-NMR)\cite{Baum}.  The MQ-NMR experiments of Baum \etal\cite{Baum} 
confirmed that for device-quality samples a characteristic feature of the microstructure 
was the presence of small clusters (of size 4--7 H atoms), and with increasing concentration of hydrogen 
these clusters merged into large clusters. Thus experimental data appear to suggest that 
the distribution is inhomogeneous at low concentration, and the microstructure mainly consists of small 
(hydrogen) clusters and a dilute distribution of hydrogen atoms dispersed in the silicon matrix. However, recent 
experimental results by Wu \etal\cite{Rutland} on hot wire chemically vapor deposited (HW-CVD) 
samples have revised this view. The authors have observed a new hydrogen distribution with 
broad line widths of about 34--39 kHz for the glow 
discharge (GD) samples of concentration 8--10\%, and about 47--53 kHz for the hot wire (HW) samples 
of concentration 2--3\%. In both the samples, the narrow line widths have been found to be of the 
order of 3--6 kHz. This implies that even at very low concentration it is possible to have  large 
clusters of hydrogen in the samples that can produce a line width as broad as 50.0 kHz.  In other words, 
the microstructure can vary significantly depending on the preparation conditions, method of deposition 
and substrate temperature, and not just on the concentration of hydrogen in the samples. 

While a great deal of information can be obtained from NMR experiments, IR 
spectroscopy~\cite{Reimer, ir, Schropp} can detect the presence of different hydride configurations 
by measuring the vibrational frequencies of hydrogen in various hydrogen bonding environments. 
The results from the IR studies by Ouwens and Schropp \etal\cite{Schropp} indicated that for 
device-quality samples 3--4\% of total hydrogen atoms resided in the network as isolated or distributed 
monohydrides (SiH). This observation is consistent with the MQ-NMR study of 
Baum \etal~\cite{Baum} on device-quality samples of hydrogenated amorphous silicon. 
The IR experiments by Manfredotti \etal\cite{ir} and Lucovsky \etal\cite{Lucovsky}, and 
the MQ-NMR study by Baum \etal\cite{Baum} provided evidence of the presence of various 
hydride configurations such as SiH, SiH$_2$ and SiH$_3$ in device-quality samples, 
whereas at high concentration it was observed that the networks could have open chains 
of (SiH$_2$)$_n$. 
The microstructure is further enriched by the presence of 
vacancies~\cite{Smets, Hoven}, voids~\cite{Smets, Mahan} and molecular hydrogen~\cite{Carlos 1980, Boyce 1985} 
in the networks. The M\"ossabauer spectroscopic studies by Hoven \etal\cite{Hoven} indicated the 
presence of vacancies in hydrogenated amorphous silicon. Recently, Smets \etal\cite{Smets} 
have studied samples by means of infrared absorption spectroscopy prepared via an expanding thermal 
plasma technique and observed that the 
microstructure is dominant by mono- and divacancy at low concentration of up to 14 at.\,\% H, and 
microvoids or voids at high concentration beyond 14 at.\,\% H. The presence of molecular hydrogen in the amorphous 
silicon network has been studied extensively by several groups using NMR~\cite{Baum, Carlos 1980, Boyce 1985}, 
infrared-absorption~\cite{Lohneysen 1984} and calorimetry experiments~\cite{Graebner 1984}. The results 
of these studies can be summarized by stating that approximately 1\% of total hydrogen can reside 
in the network in the molecular state. 

In contrast to experimental studies, there are only a handful of theoretical studies that explicitly focus 
on the hydrogen microstructure of amorphous silicon~\cite{Chak, Drabold, Raj, Lee, Kim, Zhang}. Of 
particular importance is the work by Drabold \etal\cite{Chak, Drabold} that addresses 
hydrogen-hydrogen pair separation in various dihydride configurations.  The theoretical 
studies by Kim \etal\cite{Lee,Kim} focused on the calculation of electronic structure of hydrogenated amorphous 
silicon and some properties of vacancy configurations, such as the changes in the volume of 
a vacancy upon relaxation of the network. Zhang \etal\cite{Zhang} studied the creation of higher 
order vacancies and their stability by relaxing the network from first-principles calculations. 
They found that any higher order vacancy had the tendency to evolve into a lower order vacancy 
including a monovacancy and a stable divacancy.  An unsatisfactory feature of most of the theoretical 
studies was that the vacancies or defect configurations were created by hand, which were then relaxed 
locally or globally in order to obtain a stable vacancy configuration. While such an approach 
is correct in principle and a prolonged first-principles molecular dynamics simulation can indeed 
find the minimum energy (stable) configuration of the defects, a short local or global relaxation 
may not be sufficient to eliminate the initial bias associated with the defect creation (by hand) in an 
overconstrained network. 

\section{Microstructure at low and high concentration of hydrogen}
A review of the experimental literature in the preceding section 
suggests that hydrogen can reside in various bonding environments, which typically consist of SiH, 
SiH$_2$, SiH$_3$, and the chain structure of mono- and dihydrides that are connected via Si atoms. 
Furthermore, many of these configurations can realize both in cluster and dilute phases 
including a completely isolated environment depending on the concentration of hydrogen 
in the samples. It is therefore important to study models of hydrogen-rich amorphous silicon network
both at low and high concentration.  In the following we discuss the microstructure of hydrogenated amorphous silicon for 
two representative concentrations of hydrogen. In particular, we consider two models of hydrogenated amorphous 
silicon with 7\% and 22\% hydrogen that consist of total 540 and 611 atoms respectively, and compare the results 
to the available experimental data from the literature. These models have been constructed using 
the experimentally constrained molecular relaxation (ECMR)~\cite{ecmr} technique 
developed by one of us.  The details of the construction of the models and their structural, electronic 
and vibrational properties have been studied thoroughly and published elsewhere~\cite{models}, which 
we would not repeat here.  The 540-atom model with 7\% H atoms has very few defects ($\le$ 2.9 \%), whereas the 
model with 22\% H has only a single defect. 

\begin{figure}[t]
\begin{center}
\includegraphics[width = 2.0 in,height= 2.0in , angle =0 ]{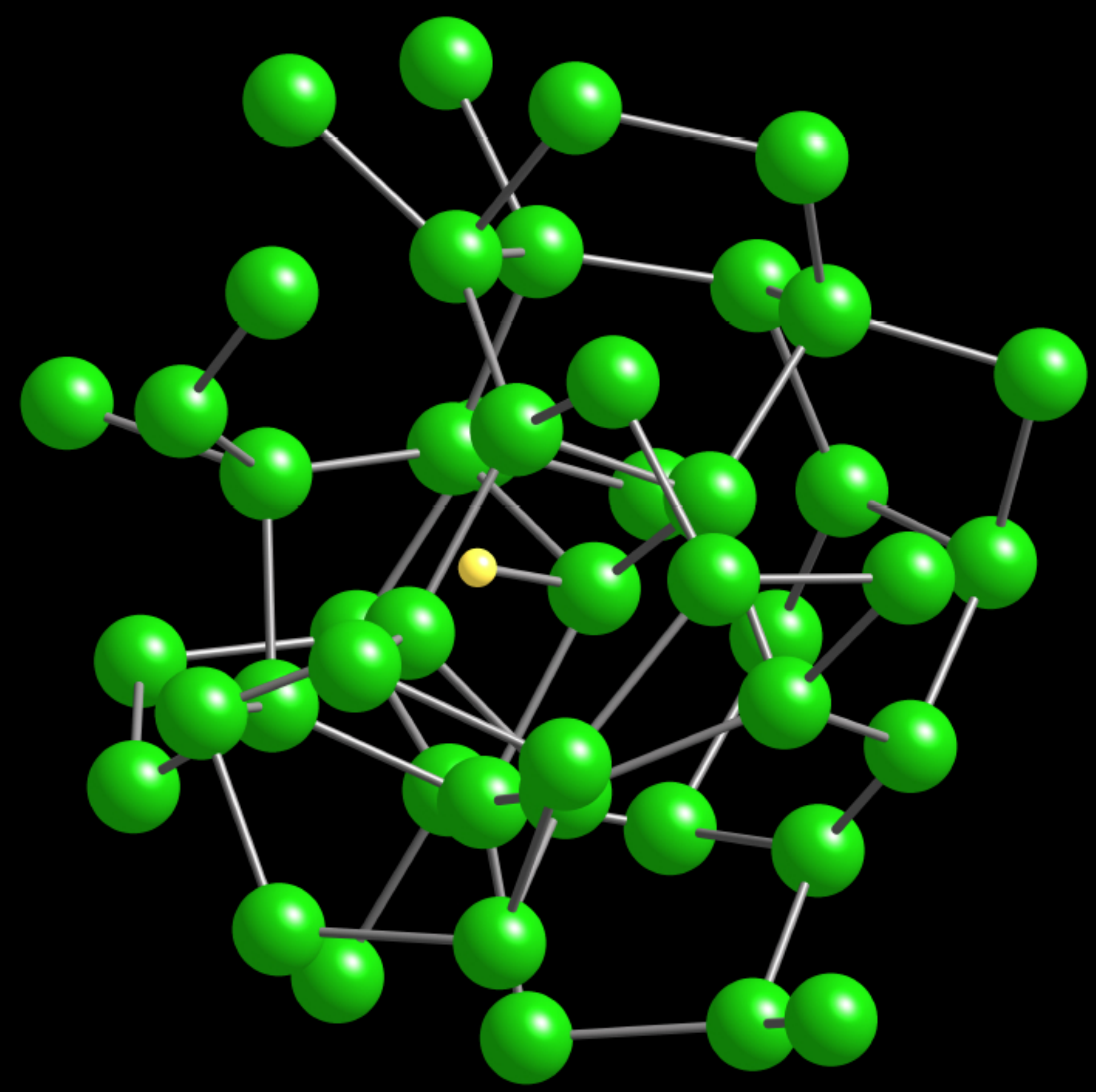}
\includegraphics[width = 2.0 in,height= 2.0in , angle =0 ]{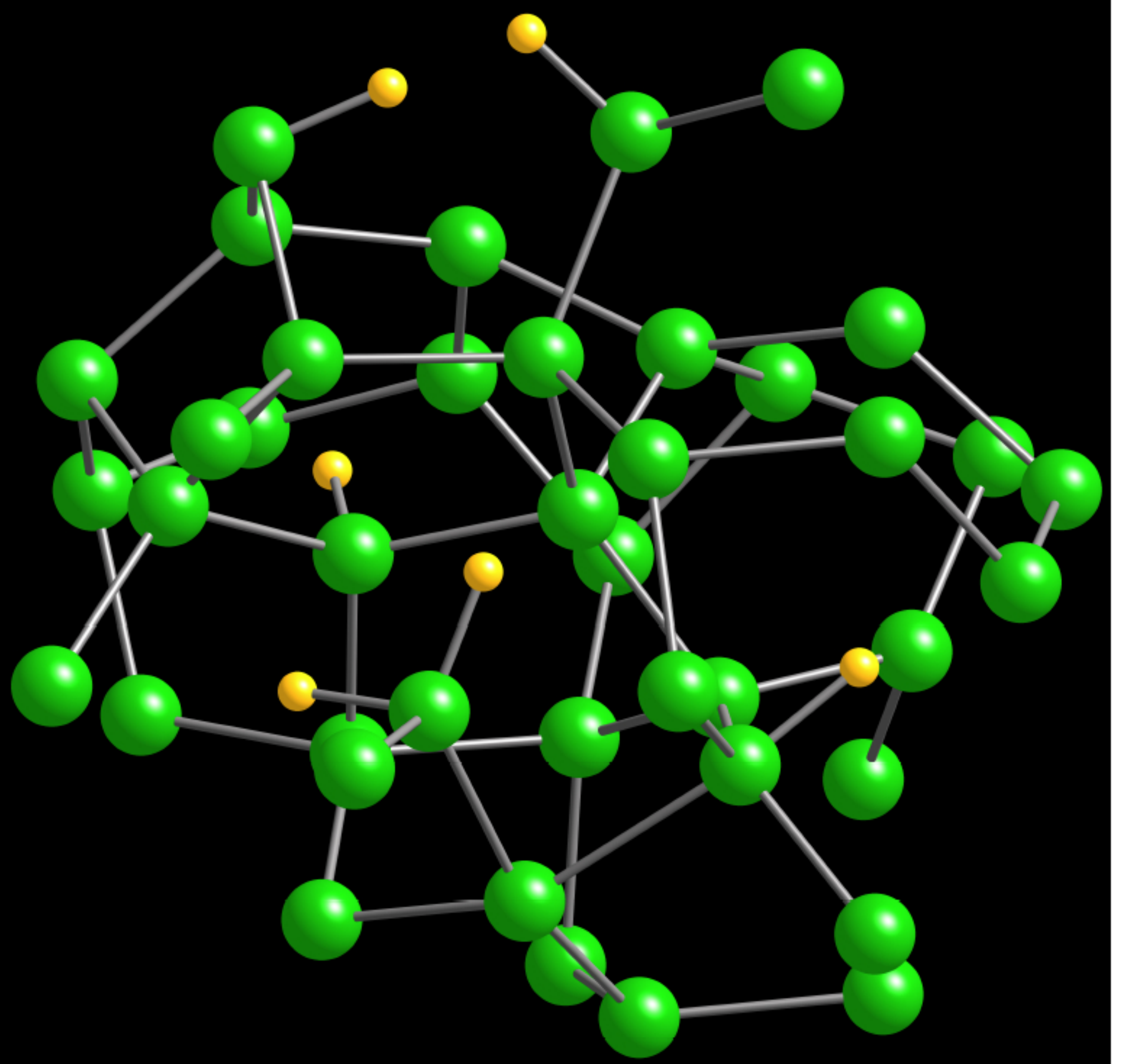}
\caption{
\label{fig1}
(Left)(a) An isolated monohydride at low concentration (7\% H) in an amorphous silicon complex of radius of 6.0\,{\AA}.  
(Right)(b) A small cluster of H atoms in a radius of 6.0\,{\AA} at low concentration.  Silicon and hydrogen atoms 
are shown in green (large) and yellow (small), respectively. 
}
\end{center}
\end{figure}

\begin{figure}[t]
\begin{center}
\includegraphics[width = 2.0 in,height= 2.0in , angle =0 ]{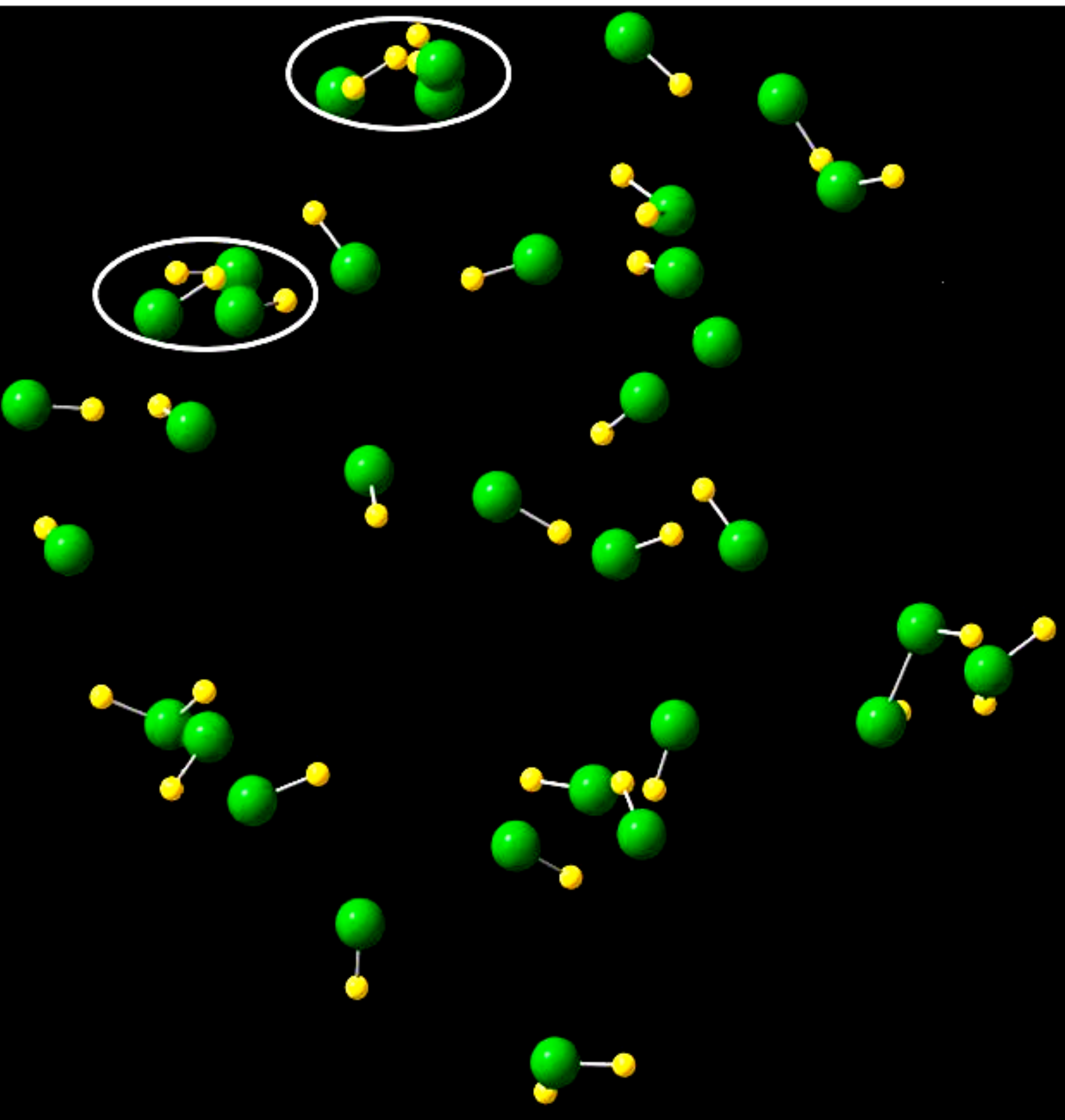}
\includegraphics[width = 2.0 in,height= 2.0in , angle =0 ]{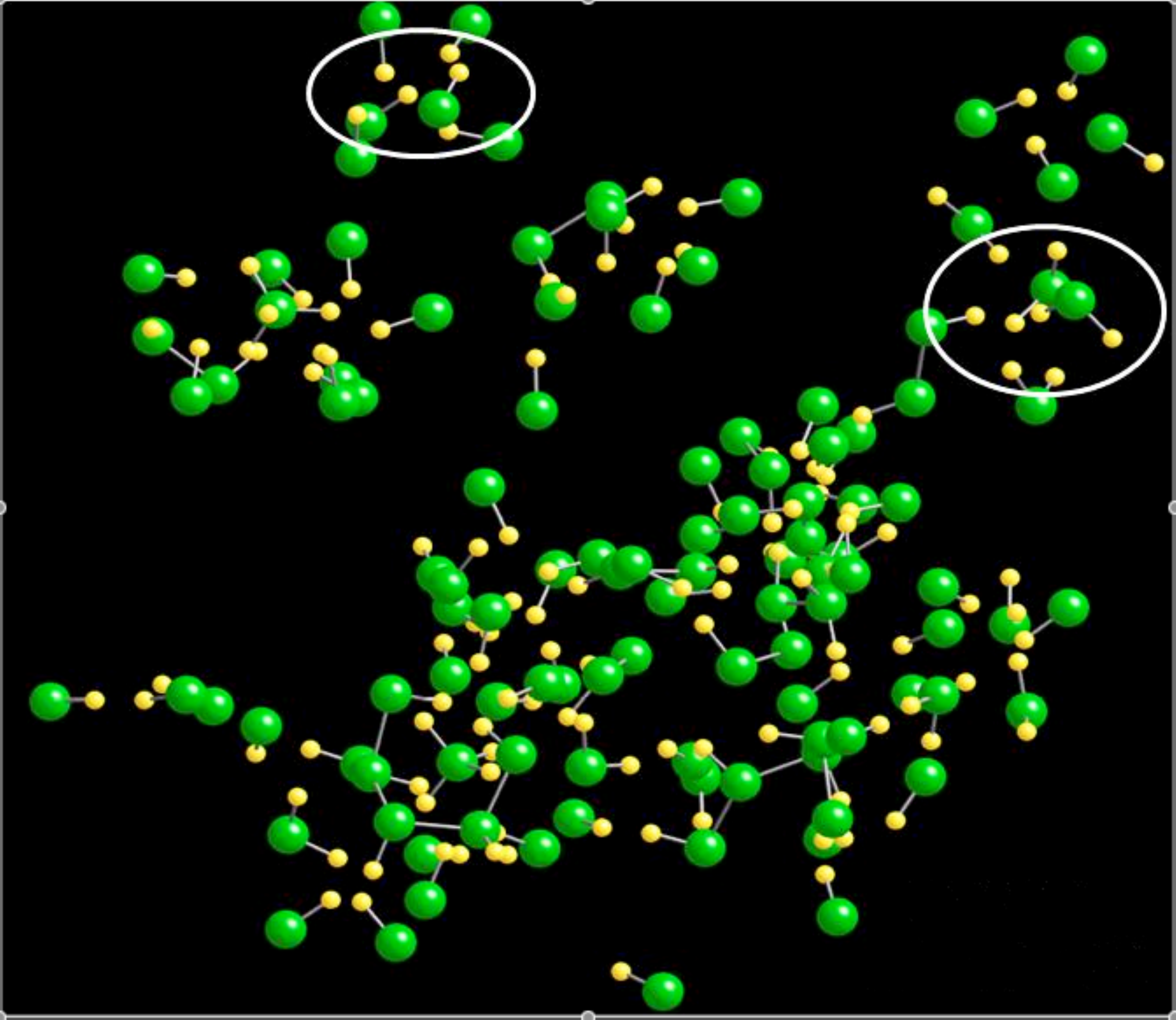}
\caption{
\label{fig2}
(Left)(a) The overall hydrogen microstructure at low concentration (7\% H) showing both clustered and dilute 
distributions of H atoms. Two small clusters are indicated in the figure.  (Right)(b) The microstructure at 
high concentration (22\% H). The various hydride configurations in clustered and sparse environments are clearly 
visible. As before, the silicon and hydrogen atoms are shown in green (large) and yellow (small), respectively. 
}
\end{center}
\end{figure}

We begin our discussion by focusing on the monohydride and the dihydride bonding configurations. The model with 
7\% H corresponds to a device-quality sample.  At this concentration, most of the 
hydrogen atoms are found to be bonded to silicon atoms as monohydrides (SiH) and dihydrides (SiH$_2$). A total 
of 75\% of all hydrogen atoms are found to reside in the form of monohydrides, whereas the remaining 
25\% are realized in dihydride configurations. An examination of the model reveals that approximately 5\% 
of the total hydrogen atoms reside as isolated monohydrides. This result is in agreement with the experiment of Ouwens and 
Schropp~\cite{Schropp}, who have observed approximately 4\% of total hydrogen atoms in the isolated phase. An 
example of such an isolated monohydride (SiH) is shown in \fref{fig1}a.  The silicon complex in the figure 
consists of about 46 Si atoms within a radius of 6.0\,{\AA} surrounding the isolated hydrogen atom. 
Apart from such an isolated monohydride configuration, many of the monohydrides and the dihydrides are realized 
in the network that are close to each other to forming a dilute distribution of hydrogen. 
The remaining hydrogen atoms form clusters that typically consist of 4--7 H atoms. 
The size of the clusters varies from 5.0\,{\AA} to 7.0\,{\AA}, and one such a cluster 
is presented in \fref{fig1}b. At low concentration all the dihydride configurations 
are realized in a clustered environment indicating the absence of isolated dihydrides in 
the model. The overall microstructure of the 540-atom (7\% H) model is shown in \fref{fig2}a 
without the silicon matrix (except for the silicon atoms bonded to hydrogen) in order to highlight 
the dilute and the densely occupied regions.  Two small clusters are encircled to indicate explicitly 
in \fref{fig2}a. The scattered distribution of such small clusters in the background of 
a dilute environment (of hydrogen) indicates that the hydrogen atoms are distributed quite 
inhomogeneously, which is consistent with the results from the nuclear magnetic resonance 
studies~\cite{Gleason, Reimer, Baum, Rutland} mentioned in section 2. 

\begin{figure}[t] 
\begin{center}
\includegraphics[width=2.0in, height=2.0in, angle =0 ]{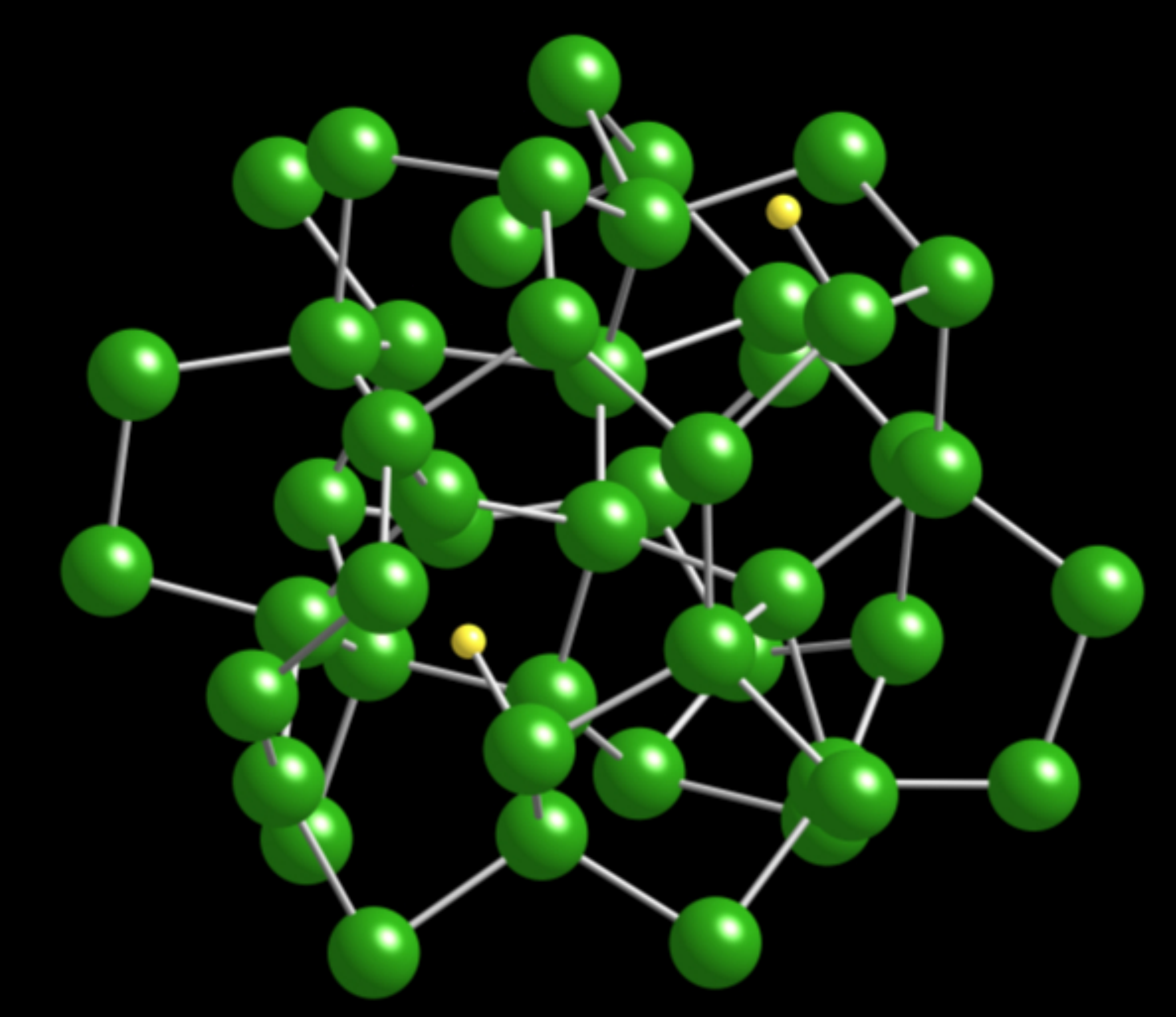}
\includegraphics[width=2.0in, height=2.0in, angle =0 ]{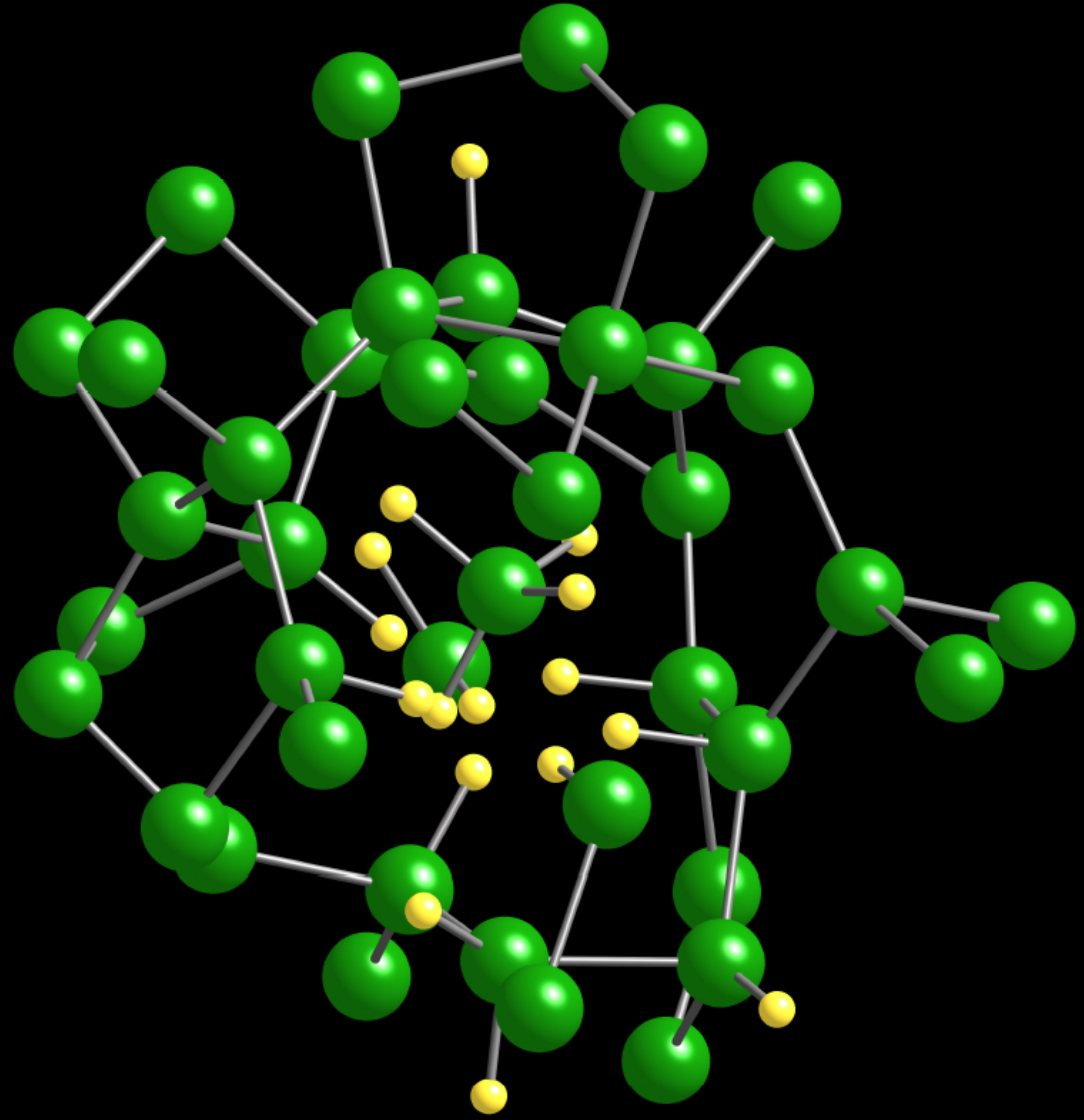}
\caption{
\label{fig3}
(Left)(a)  A pair of lone monohydrides realized in a dilute environment at high concentration (22\% H) 
in a silicon complex of linear dimension 6.0\,{\AA}. (Right)(b) A cluster of 16 H atoms in a radius 
of 6.0\,{\AA} in high concentration model (22\% H).  Silicon and hydrogen atoms are shown 
in green (large) and yellow (small) respectively. 
}
\end{center}
\end{figure}

Having discussed the model with low hydrogen content, we now address the microstructure of a model network 
with a high concentration of hydrogen.  In particular, we consider a model with 611 atoms that has total 133 
or 22\% hydrogen atoms in the network. While the concentration is somewhat higher than a typical 
device-quality sample (8--20\% H)~\cite{Carlos}, the model provides valuable insights 
about the changing nature of the microstructure with the addition of hydrogen atoms. As the hydrogen 
concentration increases, more and more hydrogen atoms are available in the network and small clusters of silicon and 
hydrogen begin to form until the concentration is high enough when large clusters appear. 
The microstructure is dominated by clusters rather than a dilute or sparse distribution 
of hydrogen at this stage. The overall distribution of hydrogen for the 611-atom model is shown in 
\fref{fig2}b. The distribution (of hydrogen) is quite inhomogeneous, which consists of both small and 
large clusters.  Two such small clusters are indicated in \fref{fig2}b that consist of 5--8 H 
atoms. A real space analysis of the model also reveals the presence of some large clusters consisting of as 
many as 16 H atoms within a radius of 6.0\,{\AA}, and is shown in \fref{fig3}b. These clusters 
mostly consist of silicon monohydrides (SiH) and dihydrides (SiH$_2$), but a few SiH$_4$ configurations 
are also found to realize in our model~\cite{Raj}. 
While the microstructure at high concentration is mostly dominated by the presence of clusters, a few 
monohydrides are also present in a very dilute or an almost isolated environment. In \fref{fig3}a we have 
shown two such lone monohydrides that are found in a silicon complex of radius 6.0\,{\AA}. 

\begin{figure}[t]
\begin{center}
\includegraphics[width=1.5 in, height=2.0 in, angle =0 ]{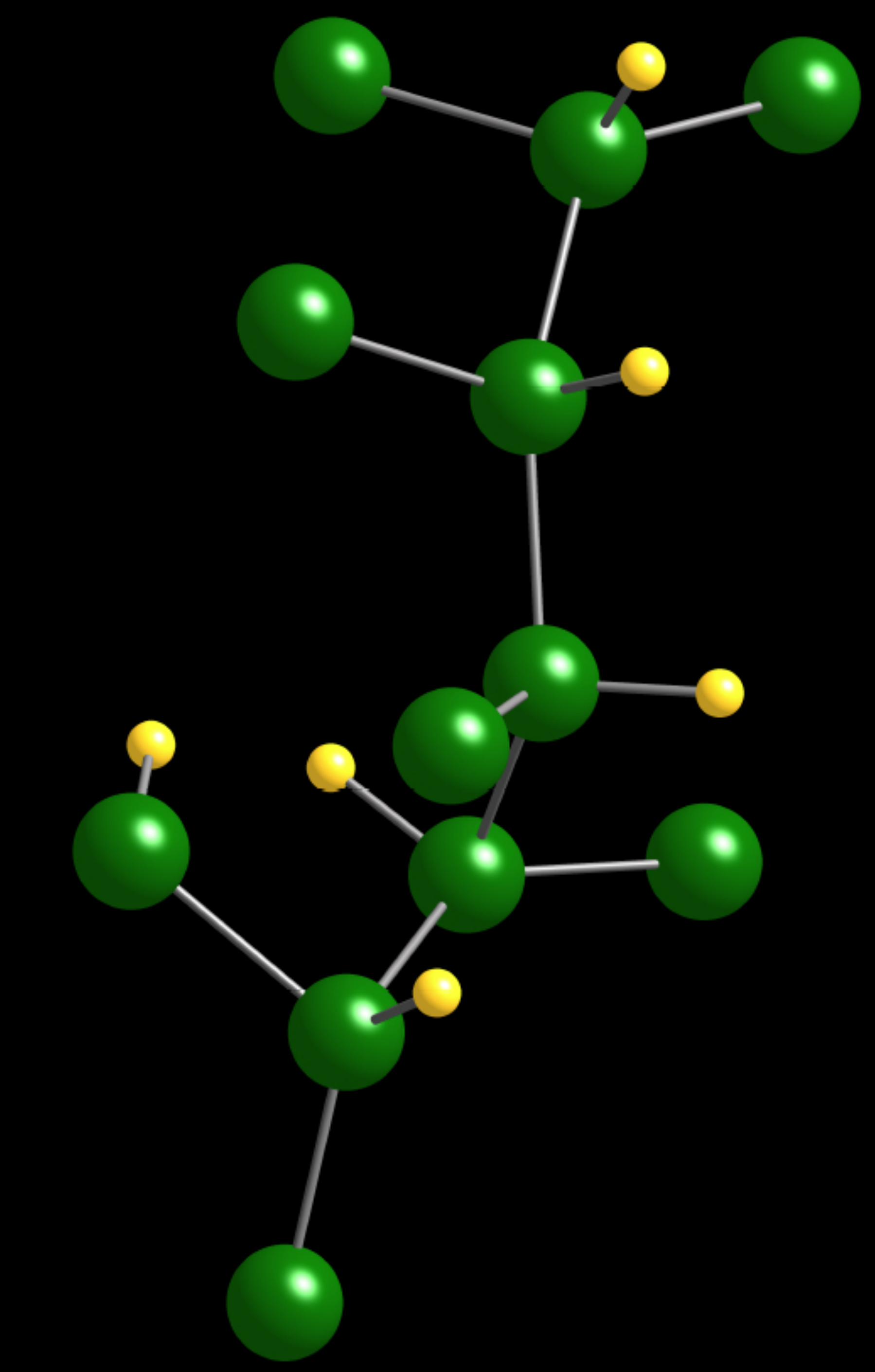}
\includegraphics[width=1.7 in, height=2.0 in, angle =0 ]{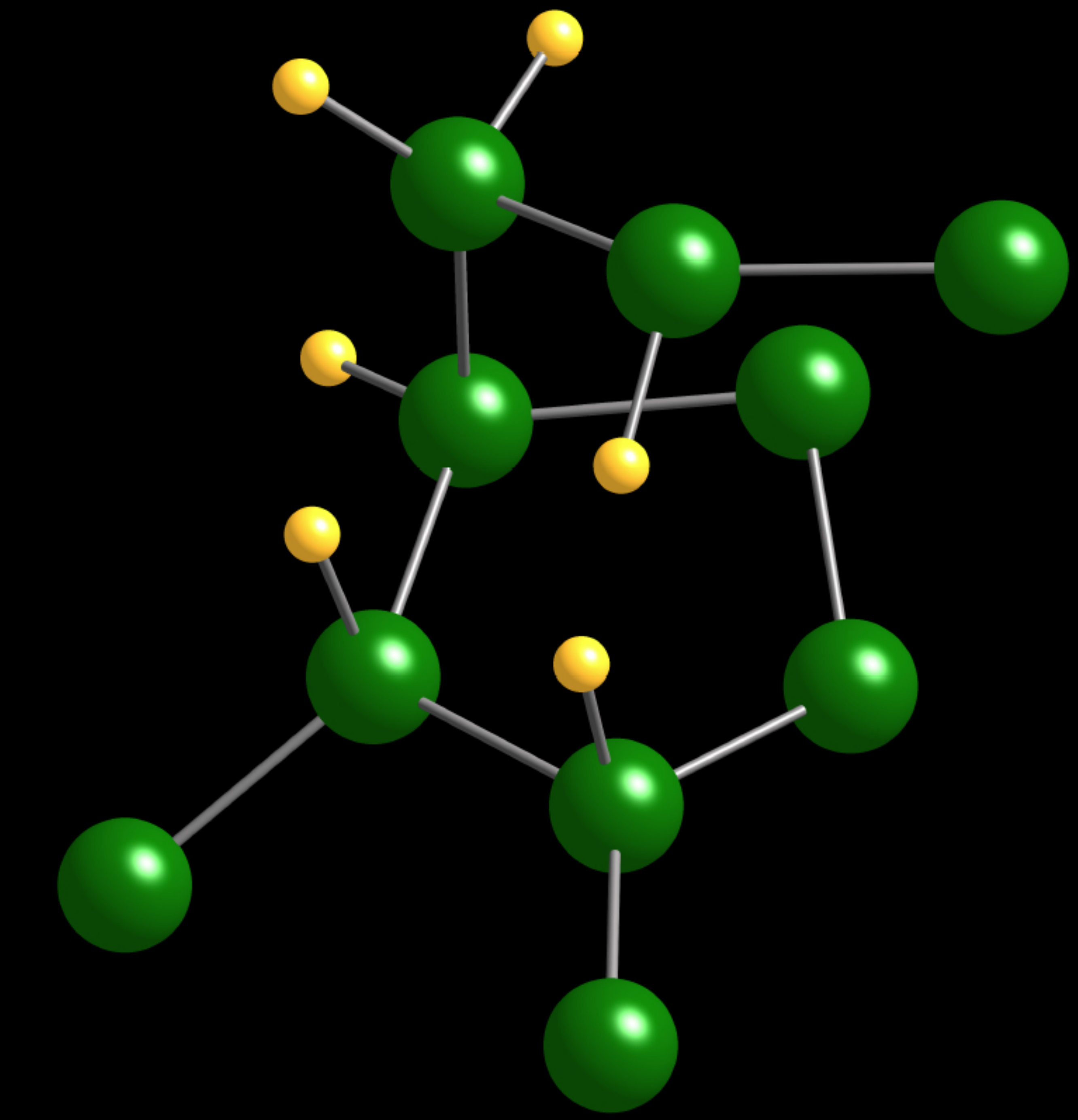}
\caption{
\label{fig4}
The formation of open chain-like structures via Si-Si bonding at high concentration (22\% H). The 
structure in the left comprises of seven monohydrides connected via silicon atoms to form an open chain. 
The figure in the right shows four monohydrides and a single dihydride. Silicon and hydrogen 
atoms are shown in green (large) and yellow (small) respectively.
}
\end{center}
\end{figure}

Finally, we should mention an important aspect of the hydrogen microstructure at high concentration. 
Some monohydrides can form an open chain-like structure at high concentration via bond formation between the 
Si atom in the monohydrides. The chain structure that we have observed in the model with 22\% hydrogen consists 
of 4 to 7 monohydrides (SiH). A few structures are also observed where monohydrides in the chain are 
replaced by dihydrides.  
Examples of such chain-like structures are shown in \fref{fig4}. 
The structure in the left (of \fref{fig4}) consists of six SiH that are connected to each other 
via Si-Si bonding, whereas the structure in the right has four monohydrides and a single dihydride. Such chain configurations 
are not observed in the model with low concentration (7\%) of hydrogen. The presence of such chain-like 
structures was reported experimentally by Lucovsky \etal\cite{Lucovsky}, Manfredotti \etal\cite{ir}, and 
Baum \etal~\cite{Baum}. 

\section{Vacancies in hydrogenated amorphous silicon} 

Unlike crystalline materials where vacancies can be uniquely defined and identified easily, the presence 
of disorder makes it difficult to do so in amorphous materials. Nonetheless, for continuous 
random networks, one can consider missing of one or two neighboring atoms in the 
topologically connected environment as a mono- or divacancy respectively. On the other hand, 
the absence of a few or many atoms in the network constitutes a microvoid or a void depending on 
the number of missing atoms. Infrared absorption spectroscopy on amorphous samples prepared via 
an expanding thermal plasma technique by Smets \etal\cite{Smets} have indicated that the 
microstructure in hydrogenated amorphous silicon samples can be characterized by the presence of 
vacancies and voids that largely depends on the amount of hydrogen present in the samples. In particular, these 
authors have observed that the microstructure is dominated by mono- and divacancy at low 
concentration of up to 14\% H, whereas microvoids or voids appear at high concentration. 
Theoretical study by Zhang \etal\cite{Zhang} also indicated the presence of mono- and divacancy 
as inherent defects in hydrogenated amorphous silicon networks. They noted that any higher 
order `$n$-vacancy' had a tendency to split into mono- or divacancy on prolonged relaxation 
of the network. In most of the theoretical studies~\cite{Lee, Kim, Zhang}, the vacancies were explicitly 
introduced by hand, and the stability of the (vacancy) configurations were studied by relaxing the 
network using an appropriate {\it ab initio} or tight-binding energy functional.

\begin{figure}[t]
\begin{center}
\includegraphics[width=2.3 in, height=2.0 in, angle =0 ]{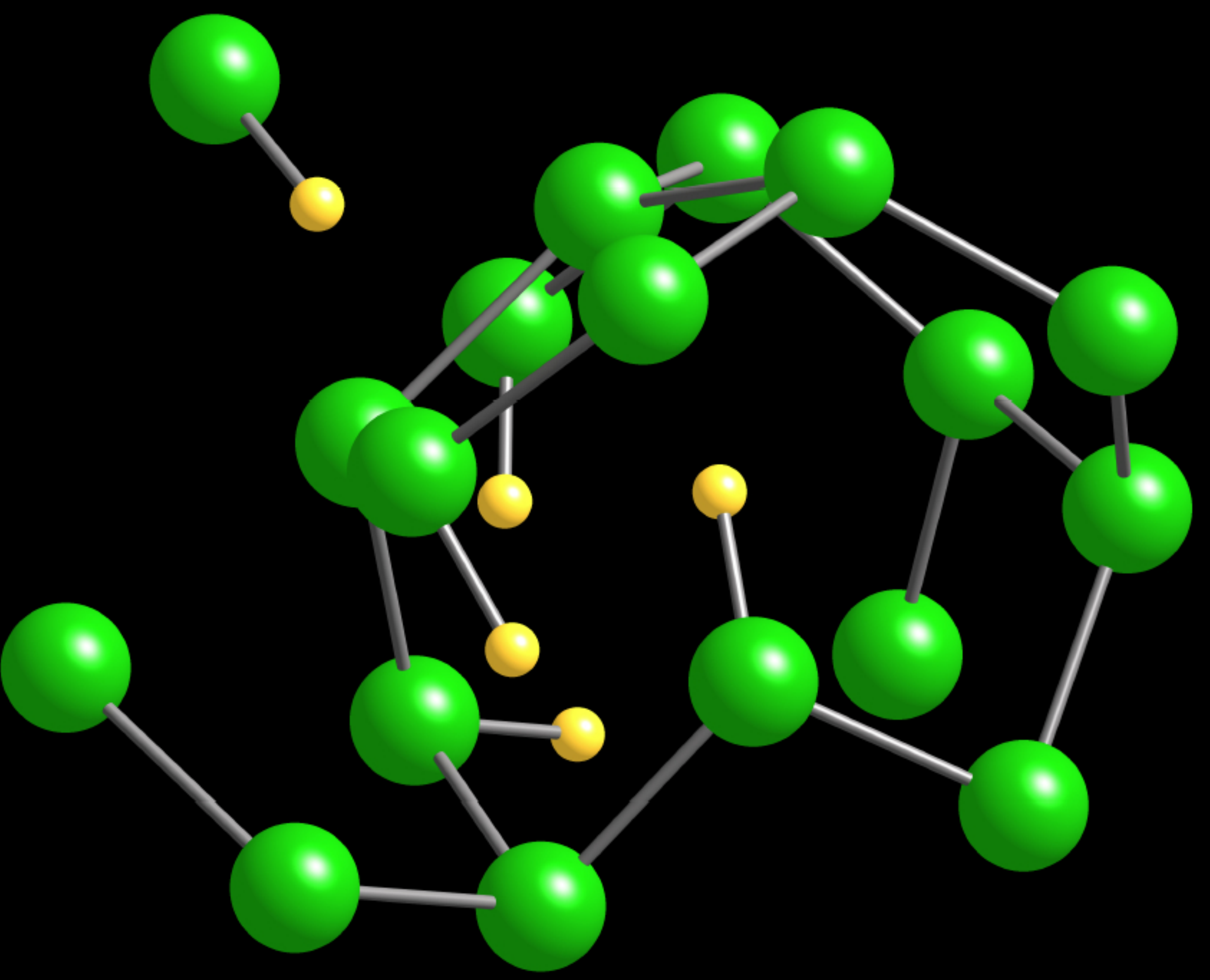}
\includegraphics[width=2.3 in, height=2.0 in, angle =0 ]{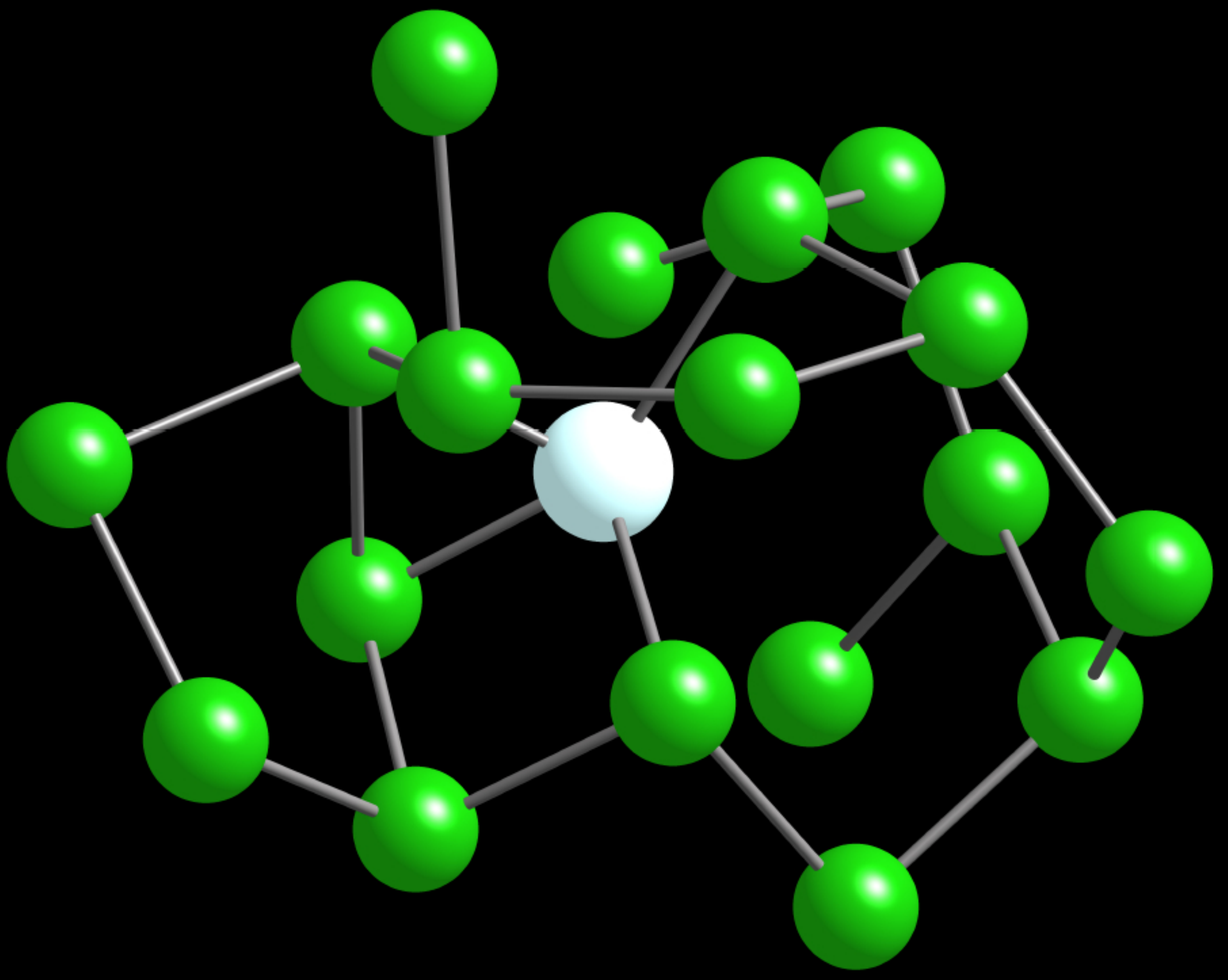}
\caption{
\label{fig5}
(Left) (a) The presence of a monovacancy (missing single atom) at low concentration (7\% H) in a region with four 
monohydrides (SiH). (Right) (b) The reconstructed network showing the missing silicon atom (in white), and how 
it would bond with the neighboring Si atoms. Silicon and hydrogen atoms are shown in green (large) and yellow (small) 
unless otherwise stated. 
}
\end{center}
\end{figure}

\begin{figure}[t]
\begin{center}
\includegraphics[width=2.0 in, height=2.0 in, angle =0 ]{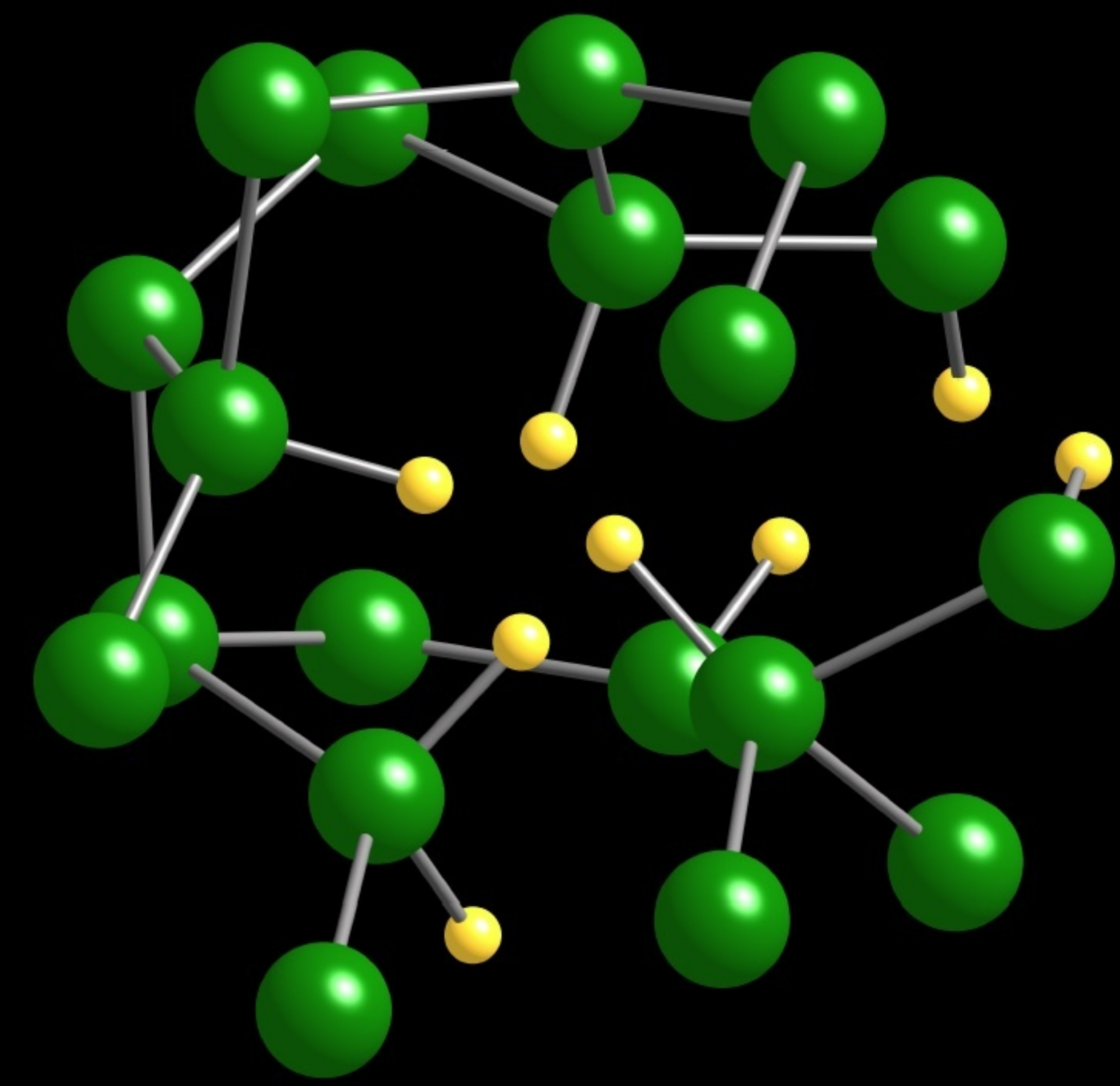}
\includegraphics[width=2.0 in, height=2.0 in, angle =0 ]{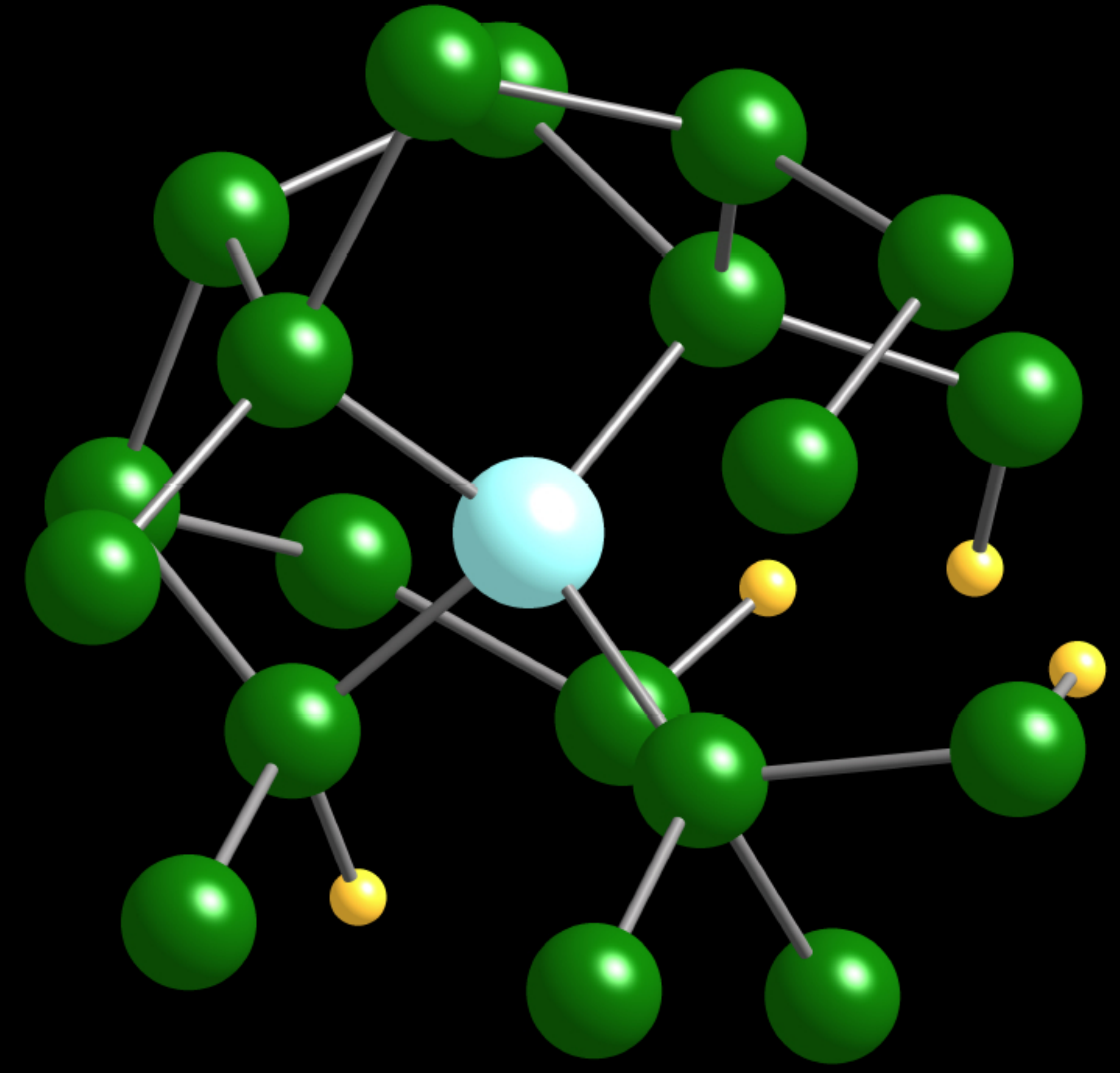}
\caption{
\label{fig6}
(Left) (a) The presence of a monovacancy (missing single atom) at high concentration (22\% H) in a region with several 
monohydrides (SiH). (Right) (b) The reconstructed figure in the right shows the missing silicon atom (white), and its bonding 
with the neighboring Si atoms. Silicon and the hydrogen atoms are shown in green (large) and yellow 
(small) except for the missing silicon atom.
}
\end{center}
\end{figure}

Motivated by the experimental results of Smets \etal\cite{Smets}, our interest lies in characterizing 
the microstructure further by searching for mono- and divacancy configurations that might be present in our 
models. A real space analysis indeed confirms that both mono- and divacancy exist in the models. This 
observation is very remarkable as vacancies are built-in or inherent 
in the networks, and are not introduced by hand as in most of the theoretical works mentioned before.
These vacancies are a characteristic feature of the models, which is a direct consequence of our unique 
model building approach via ECMR~\cite{ecmr}. In \fref{fig5}a, we have shown the region surrounding 
the monovacancy that consists of four neighboring hydrogen atoms observed in the model with 7\% H atoms.  For 
clarity of presentation and to demonstrate the origin of the monovacancy clearly, we have explicitly shown the 
missing silicon atom in the reconstructed network in \fref{fig5}b. 
The latter is constructed by removing the four hydrogen atoms, and introducing a silicon atom in the 
region. A first-principles relaxation using {\sc Siesta}~\cite{siesta} shows that 
the missing silicon atom bonds with the neighboring silicon atoms by local reconstruction of the network to 
minimize the strain. This observation also applies to the model with 22\% hydrogen, and the corresponding 
defect region is shown in \fref{fig6}a. In this case, the missing silicon atom is directly bonded to the 
silicon atoms of the neighboring monohydrides, which is clear from figures \ref{fig6}a and \ref{fig6}b.

\begin{figure}[t]
\begin{center}
\includegraphics[width=2.5 in, height=2.0 in, angle =0 ]{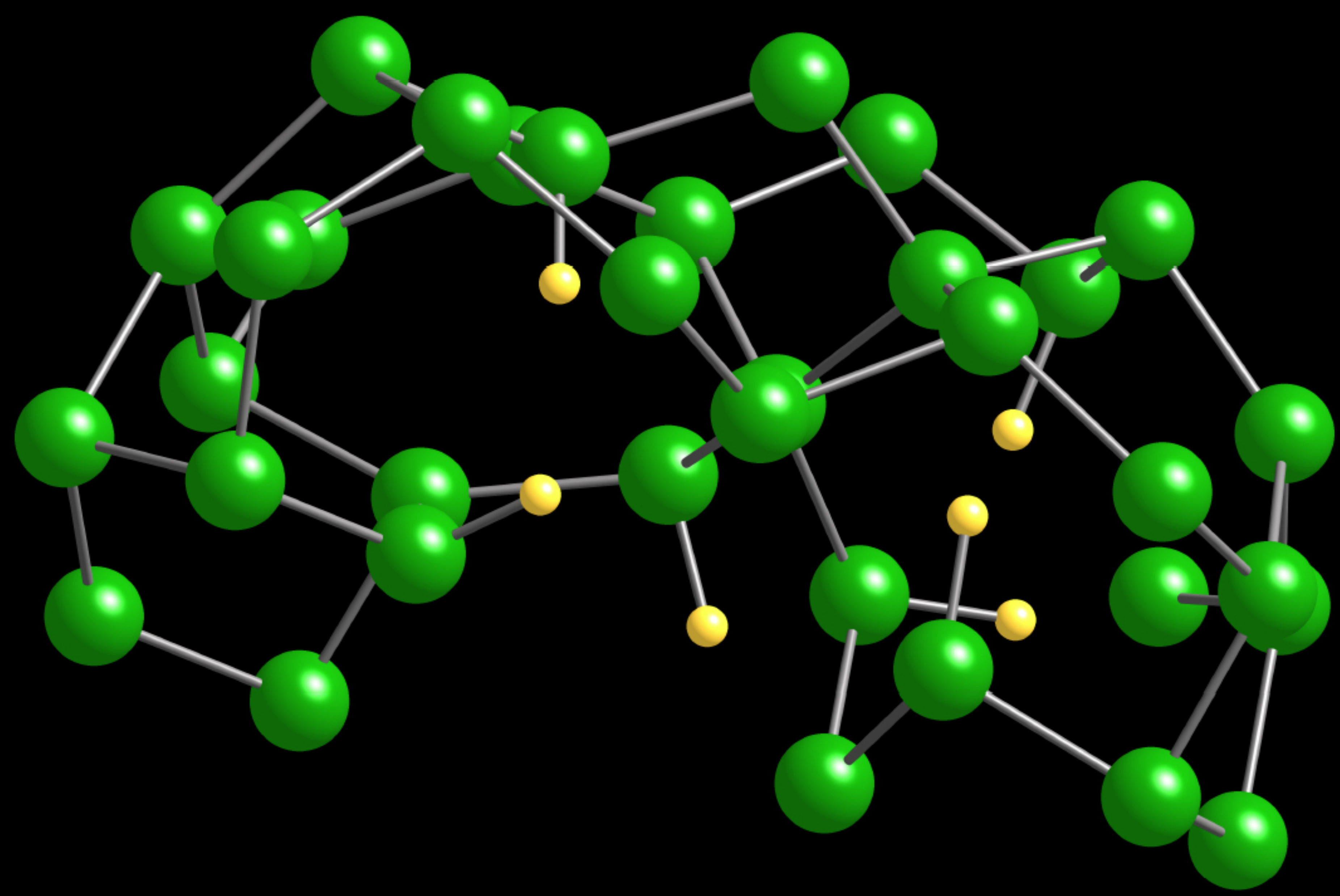}
\includegraphics[width=2.5 in, height=2.0 in, angle =0 ]{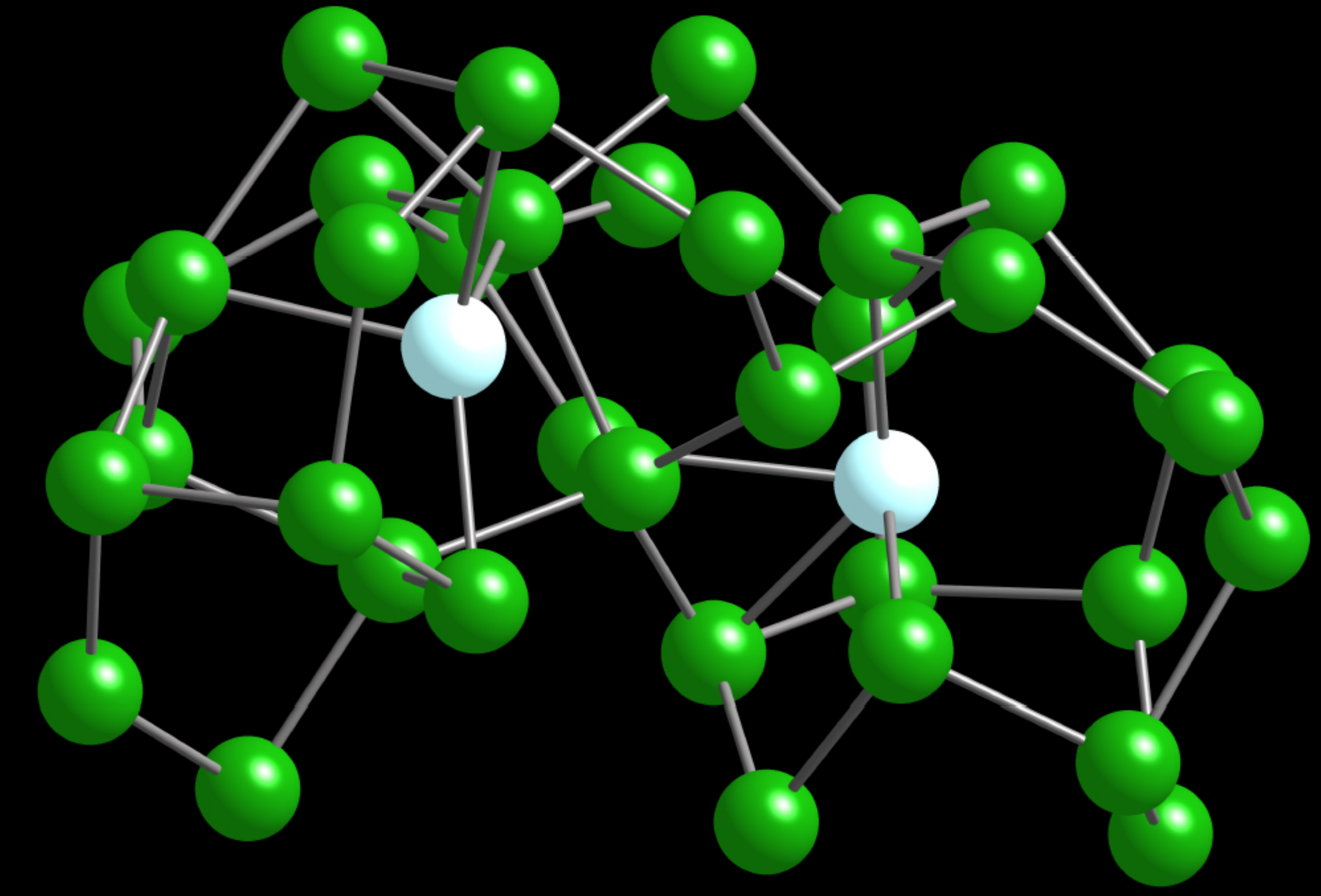}
\caption{
\label{fig7}
(Left) (a) 
The region showing a split divacancy with six monohydrides (SiH) at low concentration (7\% H). 
(Right) (b) The origin of the divacancy is illustrated by showing the missing two silicon 
atoms in the network (white). Silicon and the hydrogen atoms are painted in green (large) 
and yellow (small), respectively. 
}
\end{center}
\end{figure}

\begin{figure}[t]
\begin{center}
\includegraphics[width=2.3 in, height=2.0 in, angle =0 ]{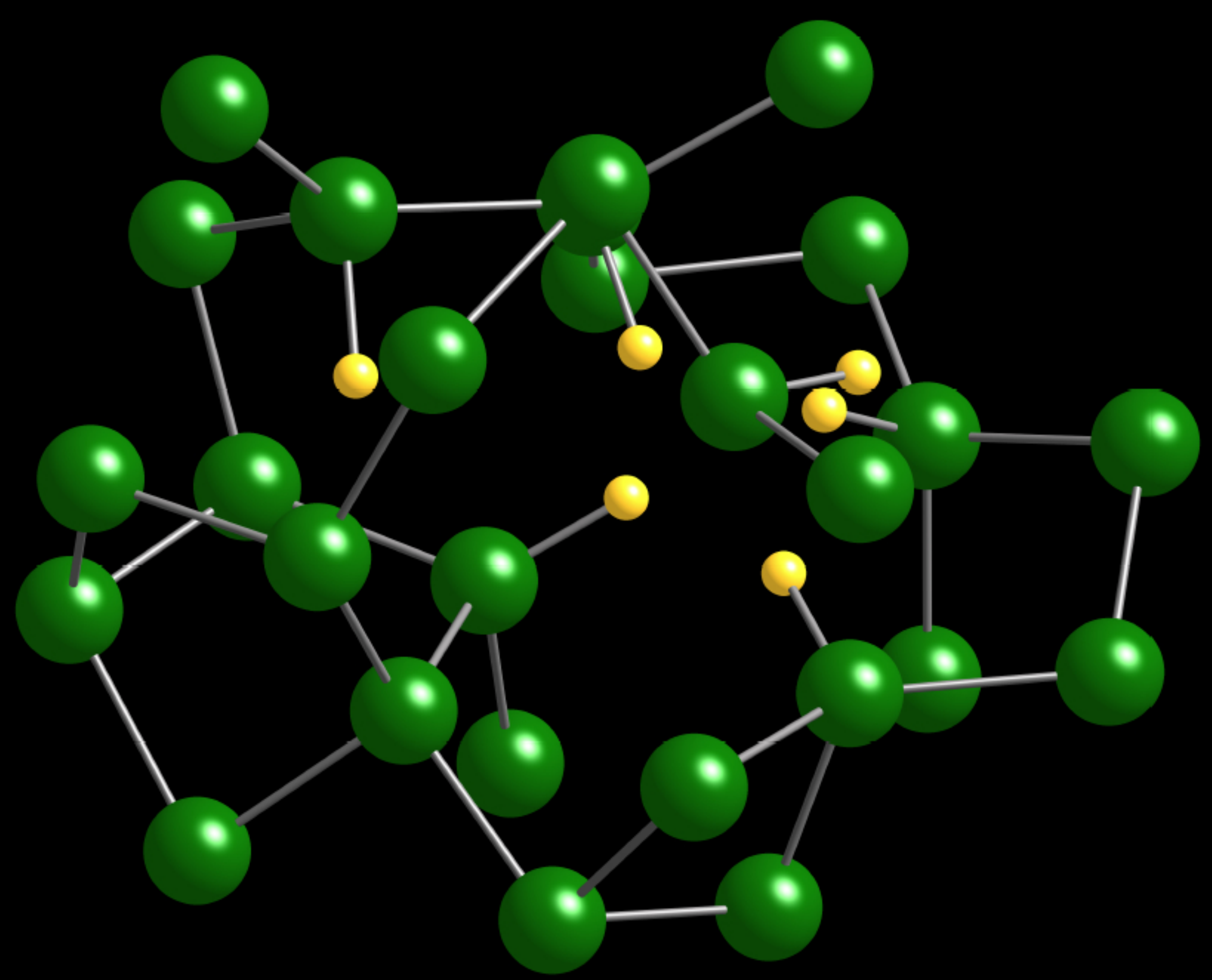}
\includegraphics[width=2.3 in, height=2.0 in, angle =0 ]{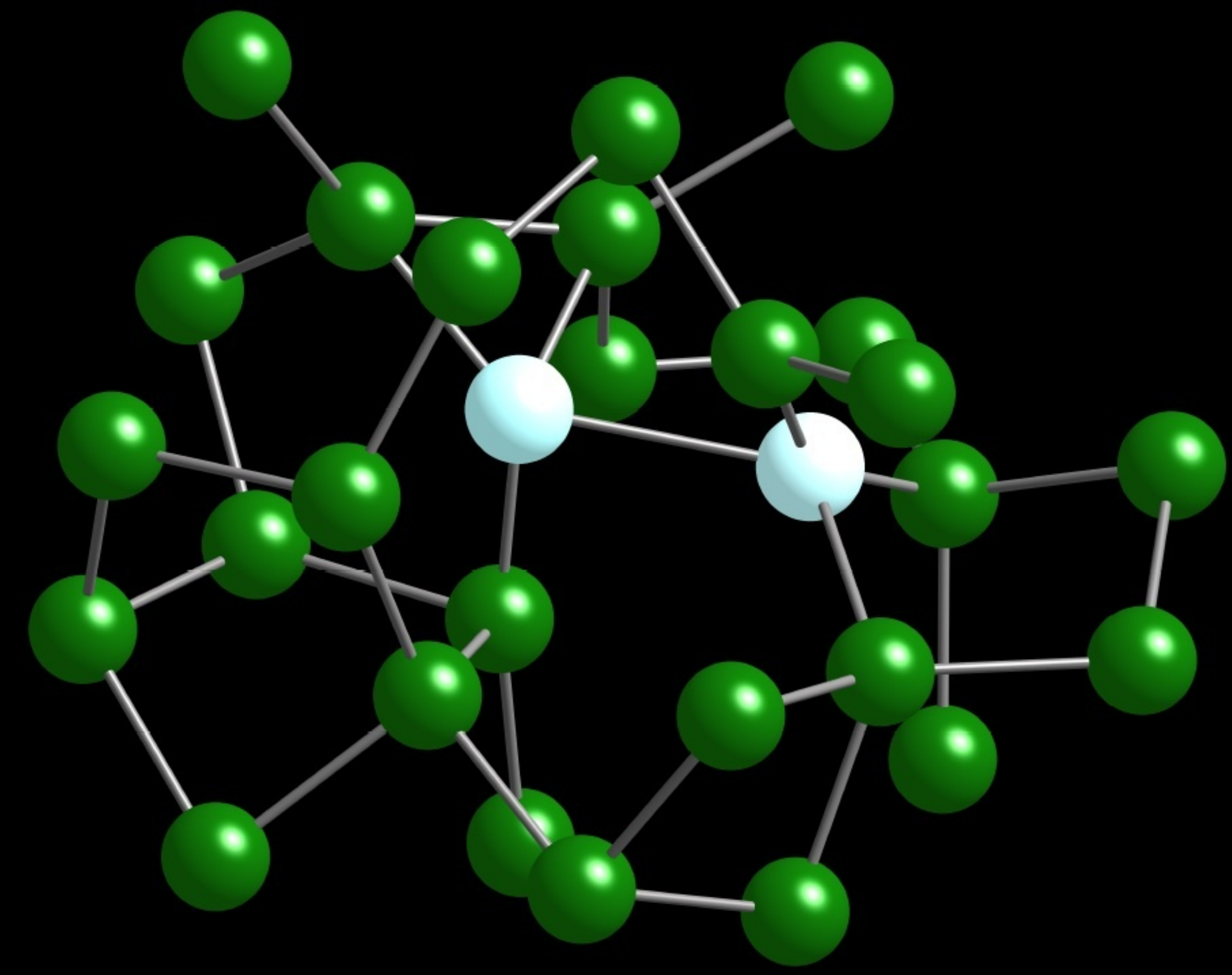}
\caption{
\label{fig8}
(Left) (a) A stable divacancy at high concentration (22\% H). (Right) (b) The reconstruction of two missing silicon 
atoms (white) as described in the text. Silicon and the hydrogen atoms are shown in green (large) and yellow 
(small) respectively. 
}
\end{center}
\end{figure}

The presence of divacancies has been also found in our models. The observed divacancies can be 
divided into two types--stable and split. The former is found in high concentration model whereas 
the latter has realized at low concentration. For the latter, the two missing silicon atoms are not 
nearest neighbor but they are in close proximity of each other typically within a distance 
of next near neighbor or so. Following Zhang~\cite{Zhang}, a split divacancy has a tendency to break 
into a stable divacancy or two monovacancies. In \fref{fig7}a, we have shown the region 
that consist of a cluster of monohydrides at low concentration (7\% H). The positions of the 
two missing silicon atoms are shown in \fref{fig7}b. The exact location is obtained by removing six 
hydrogen atoms and introducing two silicon atoms, and relaxing the resulting network via {\sc Siesta} as 
before. The missing silicon atoms are separated by a distance approximately 6.1\,{\AA}, and is 
indicative of the presence of a spilt divacancy (cf.\;\fref{fig7}a). 
The origin of the split divacancy can be understood in view of the low concentration of hydrogen in 
the network, where the microstructure is dominated by a sparse distribution of hydrogen atoms. 
At high concentration, however, stable divacancies appear. The presence of more hydrogen atoms at 
high concentration reduces the average (hydrogen) cluster-cluster separation, and thereby 
facilitates the formation of a stable divacancy. One such stable divacancy is shown in \fref{fig8}a. 
As in the earlier case, we have also indicated the two missing silicon atoms in \fref{fig8}b, which are 
nearest neighbor to each other in this case and are separated by 2.6\,{\AA}. 

In summary, the microstructure of the model networks is characterized by the presence of mono- and 
divacancy both at low and high concentration of hydrogen. A remarkable feature of the models is 
that the vacancies are built-in, and are inherent defects that have been experimentally observed 
in the concentration range studied in this work.


\section{NMR widths from Van Vleck moments: An approximate calculation}
The direct determination of the width and the shape of the NMR spectra for amorphous solids is 
a highly nontrivial task, and is difficult to address from first-principles calculations.  However, it is 
possible to calculate the first few moments of the resonance spectrum from the position of the spins
in the network. The approximate shape and the width of the resonance curve can be estimated from these 
moments, which can be compared to the values obtained from the experimental NMR data. It is, however, 
important to note that an one-dimensional NMR spectrum cannot fully represent 
every aspect of the three-dimensional spin distribution or the microstructure (of hydrogen) in the network, 
and further information is needed for a complete description of the latter. Nonetheless, the moments of an 
NMR spectrum provide valuable information such as the presence of hydrogen in isolated, sparse and in 
the clustered environment~\cite{Reimer, Baum, Rutland}. A typical NMR line spectrum can be expressed, 
in the lowest order approximation, as a combination of a Gaussian and a truncated Lorentzian 
distributions. Since the contribution to the higher moments generally comes from the wings of 
the distribution that is rarely observed in experiments, the first two or three non-zero moments should 
suffice to capture the characteristic features such as the shape and the width of the resonance 
curve~\cite{nmr-book}. 

NMR experiments on samples prepared by a variety of methods at different experimental conditions 
and concentrations reveal that at very dilute concentration randomly dispersed spins 
(via dipolar interaction) give rise to a narrow line width, whereas the presence of small clusters 
produce a broad line width in the spectrum.  These narrow and broad widths 
are generally considered to be associated with a truncated Lorentzian and a Gaussian 
broadening of the spectrum respectively, and can be constructed from theoretical arguments based on 
the moments of the resonance curve. Following Van Vleck~\cite{Vleck}, the second and the fourth moments 
of a system of $N$ particles (of spin 1/2) can be written as: 
\be
\frac{M_2}{\gamma^4 \hbar^2} = \frac{1}{2N} \sum_{j < k}^N B_{jk}^2,
\label{M2}
\ee
\bea
\frac{M_4}{\gamma^8 \hbar^4} &=& \frac{3}{16N}\sum_{ <jkl> }^N B_{jk}^2 B_{jl}^2
-\frac{1}{36N}\sum_{ <jkl>}^N B_{jk}^2(B_{jl}-B_{kl})^2 \nonumber \\ 
&+&\frac{1}{72N} \sum_{<jkl>}^N B_{jk}B_{kl}(B_{jl}-B_{jk})(B_{jl}-B_{kl}) \nonumber \\
&+& \frac{1}{8N}\sum_{j < k}^N B_{jk}^4, 
\label{M4}
\eea
where
\[
B_{ij} = \frac{3}{2}\frac{(1-3\cos^2\theta_{ij})}{r_{ij}^3}, 
\]
and the symbol $\langle jkl \rangle $ stands for no two indices being equal in the triple summation.  It should be noted that 
equations (\ref{M2}) and (\ref{M4}) above are independent of the symmetry of the system, and can be applied 
to both crystalline lattices and amorphous networks~\cite{note}. The shape of the resonance curve can be interpreted in terms 
of the second and the fourth moments. For an ideal Gaussian, $\Gamma = M_4/M_2^2 = 3$, and 
a low value of $\Gamma$ indicates a bi-modal behavior or two separated peaks in the spectrum. For a Gaussian lineshape, 
the full width at half maximum (FWHM) is given by~\cite{nmr-book}, 
\be 
\sigma_g = \sqrt{8 \, M_2 \, \ln 2}. 
\label{sigma_G} 
\ee 

At low concentration and in dilute environment the ratio $M_4/(M_2)^2$ can be large, which suggests a truncated 
Lorentzian as a useful approximation to the shape of the resonance curve.  In this case, the FWHM can be expressed 
using the second and fourth moments of the spectrum as: 
\be 
\sigma_L = \sqrt{\frac{\pi^2}{3} \frac{M_2}{\Gamma}}. 
\label{sigma_L} 
\ee

In practice, the experimental NMR spectra deviate significantly from an ideal Gaussian or a truncated Lorentzian 
behavior depending on hydrogen concentration, preparation conditions, and the degree of inhomogeneity present 
in the microstructure. Equations (\ref{sigma_G}) and (\ref{sigma_L}), therefore, provide only approximate values 
of the widths originating from the clustered and the dilute environments via dipolar interactions between the spins.  
For a small deviation (from the ideal Gaussian), one often uses the Gaussian memory function approximation and expresses
the universal line width as~\cite{Mehring}:
\be 
\sigma_u = \sqrt{\frac{2\pi\, M_2}{\Gamma - 2}} \quad \mbox{for} \quad \Gamma > 3.
\label{uni}
\ee

We now proceed to calculate the width of the NMR spectra for the model configurations with 
7\% and 22\% hydrogen atoms. The second and the fourth moments of the spectrum can be calculated using 
equations (\ref{M2}) and (\ref{M4}), respectively. The FWHM is then obtained in the Gaussian 
approximation from equation (\ref{sigma_G}). In table \ref{tab1}, we have listed the values of 
the Van Vleck moments and the overall line widths of the models in the Gaussian approximation. 
The values of $\Gamma$ for the model with high concentration suggest that the NMR spectrum  can 
be well approximated by a Gaussian lineshape, and the FWHM can be obtained from equation (\ref{sigma_G}). 
The result is consistent with the hydrogen microstructure observed in section 3.  The 
hydrogen atoms are densely distributed forming clusters with hydrogen-hydrogen separation varying from 1.6\,{\AA} 
to 2.4\,{\AA}~\cite{Raj}.  This produces a broad resonance via dipolar interaction that can be approximated 
by a Gaussian lineshape. For the model with low concentration of hydrogen  $\Gamma$ is somewhat 
larger than the ideal Gaussian value 3, which is indicative of a narrower spectrum than a Gaussian.  
The contribution from the wings of the spectrum to the higher moments is not negligible in this case, and a 
direct application of the Gaussian approximation may not provide a correct estimate of the width. 

Further difficulties can arise in characterizing the microstructure of model networks via width 
in the presence of strong inhomogeneity. The calculation of the narrow width of the resonance 
curve at high concentration via equation (\ref{sigma_L}) can be misleading in the presence of 
large hydrogen clusters.  If the contribution from a few large clusters dominates 
the moment summations, the effect of the dilute environment on the width can be missed out. 
Similarly, the presence of a single large cluster in a dilute environment (at low concentration) may not be reflected 
in the calculated value of the narrow width from model networks.  It is therefore more 
appropriate to partition the overall microstructure into regions of clustered and dilute phases, 
and to use the corresponding moments in equations (\ref{M2}) and (\ref{M4}) for calculation of 
widths for the broad and the narrow part of the spectrum.  This partition of the spins into 
clustered and dilute regions can be viewed as analogous to the deconvolution of an one-dimensional 
NMR spectrum into a broad and narrow spectrum. In the following, we discuss our results for two 
different concentrations by partitioning the microstructure into three-dimensional regions of 
dilute and clustered environments. As mentioned earlier, we can expect to see that the broad 
line width mainly originates from the clustered environment, whereas the dilute environment 
accounts for the narrow part of the spectrum. 

The results for the broad and narrow line widths from the clustered and dilute environment for each of 
the model configurations are listed in tables \ref{tab2} and \ref{tab3}. In addition to the Gaussian 
line width ($\sigma_g$) for the broad spectrum, we have also included the universal line width ($\sigma_u$) obtained from 
equation (\ref{uni}) via the Gaussian memory function approximation~\cite{Mehring}. The narrow line width of the spectrum 
is calculated assuming a truncated Lorentzian shape, and is indicated in the tables as $\sigma_L$. 
We have also included the value of the narrow line width estimated from the density of 
hydrogen atoms ($n$) in the dilute region using $\sigma_{\mbox{\small narrow}} = 4\pi\gamma^2 \hbar\,n/(3\sqrt{3})$~\cite{nmr-book}.
Since the latter is often used by experimentalists to get an approximated value of the narrow line width, this provides 
a simple way to compare our results to the experimental data. In table \ref{tab4}, we have summarized the 
experimental values of the widths obtained by different authors with the theoretical results from the models. 
The experiments of Reimer \etal\cite{Reimer} and Gleason \etal\cite{Gleason} have 
indicated that the broad line width typically ranges from 22 to 30 kHz for samples with 8 to 15\% H atoms. 
This is comparable to the values of 19.0 to 25.0 kHz obtained from the model with 7\% H. 
The calculated value of the broad line width from the model with 22\% H is somewhat larger than the 
experimental value. However, this can be understood by taking into account the presence of a large cluster 
(cf.\; \fref{fig3}b) in the model.  The presence of such large clusters is not surprising and have been 
reported by Wu \etal\cite{Rutland}, who have observed a broad line width of 50 kHz in hot-filament-assisted 
CVD deposited films. 

\begin{table}
\caption{Van Vleck moments and the overall widths for the models from the Gaussian approximation} 
\label{tab1}
\begin{indented} 
\lineup
\item[]
\begin{tabular}{@{}*{6}{l}}
    \br
    $\%$ H & Directions&$\mu_2 (\gamma^4 \hbar^2)$&$\mu_4(\gamma^8 \hbar^4)$ & $\Gamma$ & $\sigma_g$ (kHz) \\
    \mr
    7  & 100&$3.95   \times 10^{-3}$&$6.71 \times 10^{-5}$ & 4.30 & 17.8 \cr
         & 110&$3.15 \times 10^{-3}$&$5.30 \times 10^{-5}$ & 5.34 & 15.9 \cr
         & 111&$4.15 \times 10^{-3}$&$9.68 \times 10^{-5}$ & 5.64 & 18.2 \cr
    \br
    22  & 100&$3.12 \times 10^{-2}$&$2.98 \times 10^{-3}$  & 3.07  & 49.9 \cr
         & 110&$3.19 \times 10^{-2}$&$3.04 \times 10^{-3}$ & 2.99  & 50.5 \cr
         & 111&$3.26 \times 10^{-2}$&$3.18 \times 10^{-3}$ & 2.99  & 51.1 \cr
    \br
  \end{tabular}
  \end{indented}
\end{table}

\begin{table}
\caption{Deconvoluted widths for the model with 7\% H for the dilute and cluster regions } 
\label{tab2}
\begin{indented} 
\lineup
\item[]
\begin{tabular}{@{}*{5}{l}}
    \br
   Distribution &Direction & $\sigma_u$(kHz)  & $\sigma_g$ (kHz)  \cr
    \mr
   Cluster & 100 & 25.1 &  26.4    \cr
           & 110 & 22.6 &  25.4    \cr
           & 111 & 19.2 &  21.8    \cr
    \mr
   & & $\sigma_L$ (kHz)  & $\sigma_{\tt narrow}$ (kHz)   \cr
    \mr 
   Dilute  &100 & 5.3 &  2.8     \cr
           &110 & 4.9 &  2.8     \cr
           &111 & 4.9 &  2.8     \cr
    \br
\end{tabular}
\end{indented}
\end{table}

\begin{table}
\caption{Deconvoluted widths for the model with 22\% H for the dilute and cluster regions} 
\label{tab3}
\begin{indented} 
\lineup
\item[]
\begin{tabular}{@{}*{4}{l}}
    \br
   Region &Direction & $\sigma_u$(kHz)  & $\sigma_g$ (kHz)  \cr
    \mr
   Cluster & 100 & 33.8 & 40.6    \cr
           & 110 & 47.3 & 46.8   \cr
           & 111 & 44.3 & 46.4    \cr
    \mr
   & & $\sigma_L$ (kHz)  & $\sigma_{\tt narrow}$ (kHz)   \cr
    \mr 
   Dilute  &100 & 5.9 &  6.4     \cr
           &110 & 6.9 &  6.4     \cr
           &111 & 5.9 &  6.4     \cr
    \br
\end{tabular}
\end{indented}
\end{table} 
\begin{table}
\caption{Summary of experimental and theoretical values of widths}
\label{tab4} 
\begin{indented} 
\lineup
\item[]
\begin{tabular}{@{}*{4}{l}}
    \br
   Theory &($\%$ H) & $\sigma_u$ (kHz) & $\sigma_{\tt narrow}$ (kHz)  \cr
    \mr
   540-atom model & 7.0  & 19.0--25.0 & 2.8--5.3   \cr 
   611-atom model & 22.0 & 33.8--47.3 & 5.9--6.4 \cr 
    \mr 
   Experiments &($\%$H)&$\sigma_{\tt broad}$(kHz)  &$\sigma_{\tt narrow}$ (kHz)\cr
   \mr 
       Reimer \etal (1980) & 8.0--32.0 & 22.0--27.0	& 3.0--5.0  \\
       Gleason \etal (1987)& 8.0--15.0 & 25.0--30.0	& --   \\
       Wu \etal (1996)& 8.0--10.0 & 34.0--39.0	& 4.0--6.0  \\
       \qquad ''                & 2.0--3.0  & 47.0--53.0	& 3.0--6.0  \\
    \br
  \end{tabular}
  \end{indented}
  \end{table}

In summary, the broad and the narrow line widths for both the models studied here lie within the range 
of experimental values. The large value of the broad line width for high concentration model can be 
understood in terms of the presence of several large clusters with hydrogen-hydrogen separation between 
1.6\,{\AA} to 2.4\,{\AA}. Our results suggest that the shape and the width of the resonance curve 
can be approximated by a Gaussian lineshape at high concentration. However, at low concentration the shape 
begins to deviate from an ideal Gaussian toward a narrow shape, which can be approximated by a 
truncated Lorentzian distribution provided that the microstructure is dominated by dilute environment. 
However, in the presence of a strong inhomogeneous distribution of hydrogen atoms, it is difficult to quantify the 
shape of the resonance curve via single distribution, and to obtain the line width from thereof. A 
more convenient approach in this case is to describe the microstructure separately by regions of 
clustered and dilute environment, and to compute the width originating from each of these regions. 

\section{Conclusion} 
The distribution of hydrogen atoms in amorphous silicon is studied at low and high concentration 
(of hydrogen) starting from two realistic models of hydrogenated amorphous silicon.  Hydrogen 
atoms are found to be distributed in a dilute or sparse and a dense or clustered environment in the 
silicon matrix.  At low concentration, the microstructure is characterized by the presence of a few 
small clusters (4--7 H atoms) in the background of a sparse distribution of hydrogen. On the other 
hand, the microstructure is found to be strongly inhomogeneous at high concentration along with the 
presence of a few large and several small clusters.  A remarkable feature of the 
microstructure is the presence of mono- and divacancy both at low and high concentration. These 
vacancies are realized in the models as a built-in or an inherent feature and are not incorporated 
by hand during model building. The presence of such vacancies has been recently observed in experiments on hydrogenated 
amorphous silicon samples obtained via thermal plasma techniques. The widths of the NMR spectra 
of the model networks are calculated from a knowledge of the distribution of hydrogen in the 
network using the Van Vleck moments. The narrow and the broad line widths of the spectra for the 
low concentration model are found to be in the range of 3--6 kHz and 19--25 kHz, whereas the 
corresponding values for the high concentration model lie between 5.9--6.4 kHz and 34.0--48.0 kHz, 
respectively. These results are in excellent agreement with experimental data obtained from 
the NMR and the multiple-quantum NMR studies.  

\ack 
PB would like to thank the Aubrey Keith Lucas and Ella Ginn Lucas Endowment for awarding a fellowship
under the faculty excellence in research program at the University of Southern Mississippi.

\end{document}